\documentstyle[12pt]{article}
\font\tenrm=cmr10

\begin{document}
\renewenvironment{thebibliography}[1]
  { \begin{list}{\arabic{enumi}.}
    {\usecounter{enumi} \setlength{\parsep}{0pt}
     \setlength{\itemsep}{3pt} \settowidth{\labelwidth}{#1.}
     \sloppy
    }}{\end{list}}

\parindent=1.5pc

\begin{flushright} {UCRHEP-T114\\August 1993}\
\end{flushright}
\vglue 1.0cm

\begin{center}{{\bf TWO NEW SUPERSYMMETRIC OPTIONS FOR TWO\\
               \vglue 3pt
               HIGGS DOUBLETS AT THE ELECTROWEAK ENERGY SCALE\\}
\vglue 1.0cm
{ERNEST MA}\\
\baselineskip=14pt
{\it Department of Physics, University of California,}\\
\baselineskip=14pt
{\it Riverside, California 92521, USA}\\
\vglue 0.8cm
{ABSTRACT}}
\end{center}
\vglue 0.3cm
{\rightskip=3pc
 \leftskip=3pc
 \tenrm\baselineskip=12pt
 \noindent
Contrary to common belief, the requirement that supersymmetry exists and
that there are two Higgs doublets and no singlet at the electroweak
energy scale does not necessarily result in the minimal supersymmetric
standard model (MSSM).  Two interesting alternatives are presented.
\vglue 0.8cm}
{\bf\noindent 1. Introduction}
\vglue 0.4cm
\baselineskip=14pt
It is generally believed that given the gauge group SU(2) $\times$ U(1)
and the requirement of supersymmetry, the quartic scalar couplings of the
Higgs potential (consisting of two doublets and no singlet) are completely
determined in terms of the two gauge couplings.  This is actually not the
case because the SU(2) $\times$ U(1) gauge symmetry may be a remnant\cite{1}
of a larger symmetry which is broken at a higher mass scale {\underline
{together with the supersymmetry}}.  The structure of the Higgs potential
is then determined by the scalar particle content needed to precipitate
the proper spontaneous symmetry breaking and to render massive the assumed
fermionic content of the larger theory.  Furthermore, the quartic scalar
couplings are related to the gauge couplings of the larger theory
{\underline {as well as other couplings}} appearing in its superpotential.
At the electroweak energy scale, the reduced Higgs potential may contain
only two scalar doublets, but their quartic couplings may not be those of
the minimal supersymmetric standard model (MSSM).  In this talk I will
describe two explicit examples that the MSSM structure is not unique.
It is based on my very
recent work with Daniel Ng of TRIUMF,\cite{2} and with T. V. Duong.\cite{3}
\vglue 0.6cm
{\bf\noindent 2. The Two-Doublet Higgs Potential}
\vglue 0.4cm
Consider two Higgs doublets $\Phi_{1,2} = (\phi_{1,2}^+,\phi_{1,2}^0)$ and the
Higgs potential
\begin{eqnarray}
V &=& \mu_1^2 \Phi_1^\dagger \Phi_1 + \mu_2^2 \Phi_2^\dagger \Phi_2 +
\mu_{12}^2 (\Phi_1^\dagger \Phi_2 + \Phi_2^\dagger \Phi_1) \nonumber \\
&+& {1 \over 2} \lambda_1 (\Phi_1^\dagger \Phi_1)^2 + {1 \over 2} \lambda_2
(\Phi_2^\dagger \Phi_2)^2 + \lambda_3 (\Phi_1^\dagger \Phi_1) (\Phi_2^\dagger
\Phi_2) \nonumber \\ &+& \lambda_4 (\Phi_1^\dagger \Phi_2) (\Phi_2^\dagger
\Phi_1) + {1 \over 2} \lambda_5 (\Phi_1^\dagger \Phi_2)^2 + {1 \over 2}
\lambda_5^* (\Phi_2^\dagger \Phi_1)^2.
\end{eqnarray}
In the MSSM, there are the well-known constraints
\begin{equation}
\lambda_1 = \lambda_2 = {1 \over 4} (g_1^2 + g_2^2), ~~ \lambda_3 =
- {1 \over 4} g_1^2 + {1 \over 4} g_2^2, ~~ \lambda_4 = - {1 \over 2} g_2^2,
{}~~ \lambda_5 = 0,
\end{equation}
where $g_1$ and $g_2$ are the U(1) and SU(2) gauge couplings of the standard
model respectively.  Note that only the gauge couplings contribute to the
$\lambda$'s.  This is because that with only two SU(2) $\times$ U(1) Higgs
superfields, there is no cubic invariant in the superpotential and thus no
additional coupling.
\vglue 0.6cm
{\bf \noindent 3. The E$_6$-Inspired Left-Right Model}
\vglue 0.4cm
Consider now the gauge group $\rm SU(2)_L \times SU(2)_R \times U(1)$ but with
an unconventional assignment of fermions.\cite{4}  An exotic quark $h$ of
electric charge $-1/3$ is added so that $(u,d)_L$ transforms as (2,1,1/6),
$(u,h)_R$ as (1,2,1/6), whereas both $d_R$ and $h_L$ are singlets
(1,1,$-1/3$).  There are two scalar doublets $\Phi_1$ and $\chi$, as well
as a bidoublet
\begin{equation}
\eta = \left( \begin{array} {c@{\quad}c} \overline {\phi_2^0} & \eta^+ \\
- \phi_2^- & \eta^0 \end{array} \right),
\end{equation}
transforming as (2,1,1/2), (1,2,1/2), and (2,2,0) respectively.  Note that
$\Phi_1^\dagger \tilde \eta \chi$ is then an allowed term in the
superpotential, where $\tilde \eta \equiv \sigma_2 \eta^* \sigma_2$, so
that its coupling $f$ also contributes to the quartic scalar couplings of
this model's Higgs potential.

Let $G_1$ be the U(1) gauge coupling and $G_2$ the coupling of both SU(2)'s.
Then
\begin{eqnarray}
V &=& V_{soft} + {1 \over 8} (G_1^2 + G_2^2) [(\Phi_1^\dagger \Phi_1)^2 +
(\chi^\dagger \chi)^2] \nonumber \\ &+& {1 \over 4} G_2^2 [(Tr \eta^\dagger
\eta)^2 - (Tr \eta^\dagger \tilde \eta)(Tr \tilde \eta^\dagger \eta)] +
(f^2 - {1 \over 4} G_2^2) (\Phi_1^\dagger \Phi_1 + \chi^\dagger \chi)
Tr \eta^\dagger \eta \nonumber \\ &-& (f^2 - {1 \over 2} G_2^2)
(\Phi_1^\dagger \eta \eta^\dagger \Phi_1 + \chi^\dagger \eta^\dagger \eta
\chi) + (f^2 - {1 \over 4} G_1^2) (\Phi_1^\dagger \Phi_1)(\chi^\dagger \chi),
\end{eqnarray}
where $V_{soft}$ contains terms of dimensions 2 and 3, and breaks the
supersymmetry.  Let $\chi^0$ acquire a vacuum expectation value $u \neq 0$.
Then $\rm SU(2)_L \times SU(2)_R \times U(1)$ breaks down to the standard
$\rm SU(2)_L \times U(1)_Y$ with $m^2(\sqrt 2 Re \chi^0) = (G_1^2 + G_2^2)
u^2/2$ and $m^2(\eta^+,\eta^0) = G_2^2 u^2/2$.  These heavy particles can
be integrated out at the electroweak energy scale where only $\Phi_{1,2}$
are left.
\vglue 0.6cm
{\bf \noindent 4. Reduced Higgs Potential of the Left-Right Model}
\vglue 0.4cm
The quartic scalar couplings of the reduced Higgs potential at the electroweak
energy scale are now given by
\begin{eqnarray}
\lambda_1 &=& {1 \over 4} (G_1^2 + G_2^2) - {(4f^2-G_1^2)^2 \over {4(G_1^2+
G_2^2)}}, \\ \lambda_2 &=& {1 \over 2} G_2^2 - {(4f^2 - G_2^2)^2 \over
{4(G_1^2 + G_2^2)}}, \\ \lambda_3 &=& {1 \over 4} G_2^2 - {{(4f^2-G_1^2)
(4f^2-G_2^2)} \over {4(G_1^2+G_2^2)}}, \\ \lambda_4 &=& f^2 - {1 \over 2}
G_2^2, ~~ \lambda_5 = 0,
\end{eqnarray}
where the second terms on the right-hand sides of the equations for
$\lambda_{1,2,3}$ come from the cubic interactions of $\sqrt 2 Re \chi^0$.
In the limit $f=0$ and using the tree-level boundary conditions $G_2 = g_2$
and $G_1^{-2} + G_2^{-2} = g_1^{-2}$, it can easily be shown from the above
that the MSSM is recovered.  However, $f$ is in general nonzero, although
it does have an upper bound because $V$ must be bounded from below.  Hence
\begin{equation}
0 \leq f^2 \leq {1 \over 4} (g_1^2 + g_2^2) \left( 1 - {g_1^2 \over g_2^2}
\right)^{-1},
\end{equation}
where the maximum value is obtained if $V_{soft}$ is also left-right
symmetric.
\vglue 0.6cm
{\bf \noindent 5. Phenomenological Consequences}
\vglue 0.4cm
For illustration, let $f = f_{max}$ and $x \equiv \sin^2 \theta_W$, then
\begin{equation}
\lambda_1 = 0, ~~ \lambda_2 = {e^2 \over {2x}} \left[ 1 - {{2x^2} \over
{(1-x)(1-2x)}} \right] + {{g_2^2 \epsilon} \over {4M_W^2 \sin^4 \beta}},
\end{equation}
\begin{equation}
\lambda_3 = {e^2 \over {4x}} \left[ 1 - {{2x} \over {1-2x}} \right] =
- \lambda_4, ~~ \lambda_5 = 0,
\end{equation}
where
\begin{equation}
\epsilon = {{3g_2^2m_t^4} \over {8\pi^2M_W^2}} \ln \left( 1 + {\tilde m^2
\over m_t^2} \right)
\end{equation}
is an
extra term coming from radiative corrections and $\tan \beta \equiv \langle
\phi_2^0 \rangle / \langle \phi_1^0 \rangle$.  Comparing against the MSSM,
the lighter of the two neutral scalar bosons is now constrained by
\begin{equation}
m_h^2 < m_A^2 \sin^2 \beta, ~~ m_h^2 < 2 M_W^2 \left[ 1-{{2x^2} \over
{(1-x)(1-2x)}} \right] \sin^4 \beta + \epsilon,
\end{equation}
instead of $m_A^2 \cos^2 2\beta + \epsilon / \tan^2 \beta$ and
$M_Z^2 \cos^2 2\beta + \epsilon$, where $m_A$ is the mass of the pseudoscalar
boson.  Assuming $m_t$ = 150 GeV and $\tilde m$ = 1 TeV, this means that
\begin{equation}
m_h < 120~ {\rm GeV}
\end{equation}
in this model, whereas $m_h < 115$ GeV in the MSSM.  There
is also the sum rule
\begin{equation}
m_{H^\pm}^2 = m_A^2 + {1 \over 2} M_W^2 \left( 1-{{2x} \over {1-2x}} \right)
\end{equation}
instead of the corresponding
$m_{H^\pm}^2 = m_A^2 + M_W^2$ in the MSSM.  In the limit of large $m_A$,
both models reduce to the standard model with $h$ as its one Higgs boson
while keeping their respective mass upper limits.
\vglue 0.6cm
{\bf\noindent 6. The New SU(3) $\times$ U(1) Model}
\vglue 0.4cm
Another example of an extended electroweak gauge model is a new
version\cite{5,6} of $\rm SU(3) \times U(1)$.  Its salient feature is in
the choice of the electric-charge operator within SU(3).  Instead of the
usual $Q = I_3 + Y/2$, it is assumed here that $Q = I_3 + 3Y/2$.  Hence
for $\rm SU(3) \times U(1)$, we have
\begin{equation}
Q = I_3 + {3 \over 2} Y + Y',
\end{equation}
where $Y'$ is the U(1) hypercharge.  Consider now the fermionic content of
this model.  The three families of leptons transform identically as ($3^*$,
0).  Specifically, ($\ell^c, \nu_\ell, \ell)_L$ form an antitriplet with
$I_3 = (0, 1/2, -1/2)$ and $Y = (2/3, -1/3, -1/3)$.  The quarks are different:
the third family $(T,t,b)_L$ is also an antitriplet ($3^*$, 2/3), but the
first two, $(u,d,D)_L$ and $(c,s,S)_L$, are triplets (3, $-1/3$) with
$I_3 = (1/2, -1/2, 0)$ and $Y = (1/3,1/3,-2/3)$.  All the charge-conjugate
quark states are singlets.  As shown in Refs. [5] and [6], this structure
ensures the absence of all axial-vector anomalies.

The Higgs sector of this model must consist of at least three complex
triplets $(\eta^+, \eta^0, \eta^-), (\rho^0, \rho^-, \rho^{--})$, and
$(\chi^{++}, \chi^+, \chi^0)$, transforming as (3,0), (3,$-1$), and (3,1)
respectively.  At the first step of symmetry breaking, $\chi^0$ acquires
a large vacuum expectation value, so that SU(3) $\times$ U(1) breaks down
to the standard SU(2) $\times$ U(1) and the exotic quarks $D,S$ (of electric
charge $-4/3$) and $T$ (of electric charge 5/3) become massive.  The
subsequent breaking of SU(2) $\times$ U(1) is accomplished with nonzero
values of $\langle \eta^0 \rangle$ and $\langle \rho^0 \rangle$.
\vglue 0.6cm
{\bf\noindent 7. Supersymmetric SU(3) $\times$ U(1)}
\vglue 0.4cm
We now impose supersymmetry.  In addition to changing all fields to
superfields, we need to add three complex scalar superfields $(\eta'^+,
\eta'^0, \eta'^-), (\rho'^{++}, \rho'^+, \rho'^0)$, and $(\chi'^0, \chi'^-,
\chi'^{--})$, transforming as $(3^*,0), (3^*,1)$, and $(3^*,-1)$
respectively.  These are required for the cancellation of anomalies
generated by the $\rho, \eta$, and $\chi$ superfields.  The superpotential
now contains two cubic invariants $f\epsilon_{ijk} \eta_i \rho_j \chi_k$
and $f' \epsilon_{ijk} \eta'_i \rho'_j \chi'_k$ which contribute to the
Higgs potential.  The part related to the gauge interactions through
supersymmetry is given by
\begin{eqnarray}
V_D &=& {1 \over 2} G_1^2 [-\rho^*_i \rho_i + \chi^*_i \chi_i + \rho'^*_i
\rho'_i - \chi'^*_i \chi'_i]^2 \nonumber \\ &+& {1 \over 8} G_3^2 \sum_a
[\eta^*_i \lambda^a_{ij} \eta_j + \rho^*_i \lambda^a_{ij} \rho_j + \chi^*_i
\lambda^a_{ij} \chi_j \nonumber \\ &~& ~~~~~~~~
- \eta'^*_i \lambda^{*a}_{ij} \eta'_j - \rho'^*_i
\lambda^{*a}_{ij} \rho'_j - \chi'^*_i \lambda^{*a}_{ij} \chi'_j]^2,
\end{eqnarray}
where $G_1$ and $G_3$ are the U(1) and SU(3) gauge couplings respectively
and $\lambda^a_{ij}$ are the 8 conventional 3 $\times$ 3 SU(3) representation
matrices.  Similarly, the part of the Higgs potential related to the
superpotential is given by
\begin{eqnarray}
V_F &=& f^2 \sum_k [|\epsilon_{ijk} \eta_i \rho_j|^2 + |\epsilon_{ijk} \rho_i
\chi_j|^2 + |\epsilon_{ijk} \chi_i \eta_j|^2] \nonumber \\ &+& f'^2 \sum_k
[|\epsilon_{ijk} \eta'_i \rho'_j|^2 + |\epsilon_{ijk} \rho'_i \chi'_j|^2 +
|\epsilon_{ijk} \chi'_i \eta'_j|^2].
\end{eqnarray}
Let $\langle \chi^0 \rangle = u \neq 0$ and $\langle \chi'^0 \rangle = u'
\neq 0$, then the SU(3) $\times$ U(1) gauge symmetry is broken down to the
standard SU(2) $\times$ U(1).  Assume also that $\langle \eta'^0 \rangle$
and $\langle \rho'^0 \rangle$ are zero (see Ref. [3] for details) so that
it is possible to have only the doublets $\Phi_1 = (-\overline {\rho^-},
\overline {\rho^0})$ and $\Phi_2 = (\eta^+, \eta^0)$ at the electroweak
energy scale.  The parts of $V_D$ and $V_F$ which contain $\Phi_1, \Phi_2,
\chi^0$, and $\chi'^0$ are then given by
\begin{eqnarray}
V' &=& {1 \over 2} G_1^2 [(\Phi_1^\dagger \Phi_1)^2 - 2 (\Phi_1^\dagger
\Phi_1) (|\chi^0|^2 - |\chi'^0|^2) + (|\chi^0|^2 - |\chi'^0|^2)^2]
\nonumber \\ &+& {1 \over 6} G_3^2 [ (\Phi_1^\dagger \Phi_1 + \Phi_2^\dagger
\Phi_2)^2 - 3 (\Phi_1^\dagger \Phi_2) (\Phi_2^\dagger \Phi_1) \nonumber \\
&~& ~~~~~ - (\Phi_1^\dagger \Phi_1 + \Phi_2^\dagger \Phi_2) (|\chi^0|^2
- |\chi'^0|^2) + (|\chi^0|^2 - |\chi'^0|^2)^2] \nonumber \\ &+& f^2
[(\Phi_1^\dagger \Phi_2) (\Phi_2^\dagger \Phi_1) + (\Phi_1^\dagger \Phi_1
+ \Phi_2^\dagger \Phi_2) |\chi^0|^2].
\end{eqnarray}
\vglue 0.6cm
{\bf\noindent 8. Reduced Higgs Potential of the SU(3) $\times$ U(1) Model}
\vglue 0.4cm
Since $\langle \chi^0 \rangle = u$ and $\langle \chi'^0 \rangle = u'$, there
are cubic interactions in $V'$ involving $\chi^0$ and $\Phi_{1,2}$ as well
as $\chi'^0$ and $\Phi_{1,2}$.  These have to be taken into account in
obtaining the effective quartic scalar couplings $\lambda_i$ of Eq. (1).
However, because $\sqrt 2 Re \chi^0$ and $\sqrt 2 Re \chi'^0$ are not mass
eigenstates, we need to consider their 2 $\times$ 2 mass-squared matrix
given by
\begin{equation}
{\cal M}^2 = \left( \begin{array} {c@{\quad}c} M^2 \cos^2 \gamma + M'^2
\sin^2 \gamma & -(M^2 + M'^2) \sin \gamma \cos \gamma \\ -(M^2 + M'^2)
\sin \gamma \cos \gamma & M^2 \sin^2 \gamma + M'^2 \cos^2 \gamma \end{array}
\right),
\end{equation}
where $M^2 = 2 (G_1^2 + G_3^2/3) (u^2 + u'^2), \tan \gamma \equiv u'/u$,
and $M'$ is the mass of the heavy pseudoscalar boson $\sqrt 2 (\sin \gamma
Im \chi^0 - \cos \gamma Im \chi'^0)$ which has no cubic coupling to
$\Phi_{1,2}$.  The determinant of ${\cal M}^2$ is equal to $M^2 M'^2 \cos^2
2 \gamma$.  Hence
\begin{eqnarray}
\lambda_1 &=& {1 \over 3} G_3^2 + G_1^2 - {{2(u^2 + u'^2)} \over {M^2 M'^2
\cos^2 2 \gamma}} [(f^2 - G_1^2 -G_3^2/6)^2 \cos^2 \gamma ({\cal M}^2)_{22}
\nonumber \\ &~& ~~~~~~~~ -2(f^2 - G_1^2 - G_3^2/6) (G_1^2 + G_3^2/6) \sin
\gamma \cos \gamma ({\cal M}^2)_{12} \nonumber \\ &~& ~~~~~~~~ + (G_1^2 +
G_3^2/6)^2 \sin^2 \gamma ({\cal M}^2)_{11}],
\end{eqnarray}
and so forth.

In the limit $f = 0$,
\begin{equation}
\lambda_1 = \lambda_2 = {{G_3^2 (G_3^2 + 4 G_1^2)} \over {4 (G_3^2 +
3 G_1^2)}}, ~~ \lambda_3 = {{G_3^2 (G_3^2 + 2 G_1^2)} \over {4 (G_3^2 +
3 G_1^2)}}, ~~ \lambda_4 = -{1 \over 2} G_3^2, ~~ \lambda_5 = 0.
\end{equation}
Assuming the tree-level relations $g_2 = G_3$ and $g_1^{-2} = G_1^{-2}
+ 3 G_3^{-2}$, we then have $G_1^2 = g_1^2 g_2^2/ (g_2^2 - 3 g_1^2)$ and
the MSSM conditions, {\it i.e.} Eq. (2), are obtained as expected.  Since
$f \neq 0$ in the general case, the Higgs potential of this model differs
from that of the MSSM even though there are only two Higgs doublets at the
electroweak energy scale.  It also differs from that of the left-right
model discussed already.  The $f^2$ and $f^4$ terms in $\lambda_{1,2,3}$
depend on $\gamma$ and the $f^4$ terms on $M^2/M'^2$ as well.  For
illustration, let us take the special case $\cos \gamma = 1$, then
\begin{eqnarray}
\lambda_1 &=& {1 \over 4} (g_1^2 + g_2^2) + f^2 \left( 1 + {{3g_1^2} \over
g_2^2} \right) - {{3f^4} \over g_2^2} \left( 1 - {{3g_1^2} \over g_2^2}
\right), \\ \lambda_2 &=& {1 \over 4} (g_1^2 + g_2^2) + f^2 \left( 1 -
{{3g_1^2} \over g_2^2} \right) - {{3f^4} \over g_2^2} \left( 1 - {{3g_1^2}
\over g_2^2} \right), \\ \lambda_3 &=& - {1 \over 4} g_1^2 + {1 \over 4}
g_2^2 + f^2 - {{3f^4} \over g_2^2} \left( 1 - {{3g_1^2} \over g_2^2}
\right), \\ \lambda_4 &=& - {1 \over 2} g_2^2 + f^2, ~~~ \lambda_5 = 0.
\end{eqnarray}
The requirement that $V$ be bounded from below puts an upper bound on $f^2$
so that
\begin{equation}
0 \leq f^2 \leq {1 \over 2} g_2^2.
\end{equation}
\vglue 0.4cm
{\bf\noindent 9. Phenomenological Consequences}
\vglue 0.4cm
Let us now specialize further to the case $f = f_{max}$, we then find
\begin{equation}
\lambda_1 = 4g_1^2, ~~ \lambda_2 = g_1^2, ~~ \lambda_3 = 2g_1^2, ~~
\lambda_4 = \lambda_5 = 0.
\end{equation}
The equality of $\lambda_4$ and $\lambda_5$ means that an accidental
custodial SU(2) symmetry exists\cite{1} so that the charged Higgs boson
$H^\pm$ and the pseudoscalar Higgs boson $A$ form a triplet with a common
mass given by
\begin{equation}
m_A^2 = {{-2\mu_{12}^2} \over {\sin 2 \beta}}.
\end{equation}
The lighter of the two neutral scalar bosons is now constrained by
\begin{equation}
m_h^2 \leq 4 M_Z^2 \sin^2 \theta_W (1 + \cos^2 \beta)^2 + \epsilon,
\end{equation}
where $\epsilon$ is given by Eq. (12), as well as
\begin{equation}
m_h^2 \leq {{m_A^2 (1 + \cos^2 \beta)^2 + 4 \epsilon \cot^2 \beta} \over
{1 + 3 \cos^2 \beta}}.
\end{equation}
Hence $m_h$ has an upper bound of $4 M_Z \sin \theta_W$ at tree level and
it goes up to about 189 GeV after radiative corrections assuming $m_t =
150$ GeV and $\tilde {m} = 1$ TeV.
\vglue 0.6cm
{\bf\noindent 10. Outlook}
\vglue 0.4cm
If supersymmetry exists and future experiments discover two and only two
Higgs doublets at the electroweak energy scale, it does not mean
necessarily that the MSSM will be confirmed.  If either of the above models
with $f \neq 0$ is found, then it will point to a larger theory, {\it i.e.}
$\rm SU(2)_L \times SU(2)_R \times U(1)$ or $\rm SU(3) \times U(1)$
at a higher energy scale.
\vglue 0.4cm
{\bf\noindent 11. Acknowledgement}
\vglue 0.4cm
I thank Profs. Jewan Kim and Jihn E. Kim and other organizers of the 14th
International Workshop on Weak Interactions and Neutrinos for their
great hospitality and a very stimulating meeting.
This work was supported in part by the U. S. Department of Energy under
Contract No. DE-AT03-87ER40327.
\vglue 0.6cm
{\bf\noindent 12. References \hfil}
\vglue 0.4cm

\vglue 0.6cm
{\bf\noindent 13. Appendix: The Usual Singlet Case}
\vglue 0.4cm
The $\chi^0$ and $\chi'^0$ superfields discussed in this talk are SU(2)
$\times$ U(1) singlets, but they are different from the usual singlet
superfield that is sometimes added to the MSSM.  The latter must actually
also be a singlet under any nontrivial larger symmetry containing SU(2)
$\times$ U(1).  This fact is not generally appreciated.

In the usual singlet case, the superpotential is given by
\begin{equation}
W = \mu H_1 H_2 + h H_1 H_2 N + {\rm terms~containing~only}~N.
\end{equation}
If $N$ is heavy but $\langle N \rangle = 0$, then $\lambda_4 = h^2 - g_2^2/2$
in analogy with Eqs. (8) and (26), but the other $\lambda_i$'s are as in
the MSSM.

\end{document}